\documentclass[a4paper,12pt]{article}
\usepackage{amsmath}
\usepackage{amsfonts}
\usepackage{amssymb}

\title{{\textsc{\begin{LARGE}Understanding Gravity:\end{LARGE}}\\Some Extra Dimensional Perspectives}}
\author{{$\hbox{V H Satheeshkumar}^{1,2 \dagger }$} and {{$\hbox{P K Suresh}^{1 *}$}} \\ 
$^1$ \begin{small}School of Physics, University of Hyderabad, Hyderabad 500 046, India. \end{small} \\ 
$^2$ \begin{small}Department of Physics, Baylor University, Waco, Texas, 76798-7316, U.S.A. \end{small} \\ 
$^{\dagger \dagger}$ {\begin{small}\texttt{VH{\textunderscore}Satheeshkumar@baylor.edu}\end{small}}\\
 $^*$ \begin{small}\texttt{pkssp@uohyd.ernet.in}\end{small}}

\date{}
\begin{document}

\maketitle

\begin{abstract}
Gravity is one of the most inexplicable forces of nature, controlling everything, from the expansion of the Universe to the ebb and flow of ocean tides. The search for the laws of motion and gravitation began more than two thousand years ago but still we do not have the complete picture of it. In this article, we have outlined how our understanding of gravity is changing drastically with time and how the previous explanations have shaped the most recent developments in the field like superstrings and braneworlds. 
\end{abstract}
\newpage
\section{Intoduction}

Gravity is an immediate fact of everyday experience, but its fundamental understanding presents some of the deepest theoretical and experimental challenges in physics today. Gravitational physics is concerned with some of the most exotic large scale phenomena in the universe. But it is also concerned with the microscopic quantum structure of spacetime and the unification of all fundamental forces of nature. Gravity is thus important on both the largest and smallest scales considered in contemporary physics and remains one of the greatest challenges of twenty-first century science. 

Gravity dominates the large-scale structure of the universe only by default. Matter arranges itself to cancel electromagnetism, and the strong and weak forces are intrinsically short range. At a more fundamental level, gravity is extravagantly feeble. Where does this outlandish disparity come from? 
Perhaps the most unusual thing about gravity we know about is that, unlike the other forces of nature, gravity is intimately related to space and time. Why is it so different from the other interactions? Why has it not been able to unify it with the rest? Some attempts to understand quantum gravity have required that we live in more than four dimensions! If so, why do we not see the other dimensions? How are these extra dimensions hidden from our world? Is there a way to detect them? It is the aim of this article as to provide a short summary of the present status of these extra-dimensional theories of gravitation. But before graduating to extra dimensional theories, we will have a look at the well established theories of gravity due to Newton and Einstein. 

\subsection{Newton's Gravity}

The very earliest ideas regarding gravity must have been based on every day experience. For example, objects fall unless they are supported and climbing a hill is harder than walking on a level. Aristotle was the first to give some reasoning for these observed facts. In his view, the whole universe is made up of four concentric spheres, the innermost being the Earth, then comes Water, Air and Fire. Since stone is more of the ``Earth'' type, it falls down on the Earth when thrown up! Ptolemy extended this view to the heavens and ended up with the geocentric theory. Aristotle's notion of the motion impeded understanding of gravitation for a long time. Copernicus's view of the solar system was important as it allowed sensible consideration of gravitation. Kepler's laws of planetary motion were based on the volumes of observational data collected by Tycho Brahe. Galileo's understanding of the motion and falling bodies  was through his inclined plane experiments. But till then nobody knew what is the `cause' for such a motion. This set the scene for Isaac Newton's theory of gravity which was presented in his treatise The Principia \cite{Newton} in 1687.

Newton started from Galileo's law of falling objects and applied it to an unlikely object, the Moon which seems to flout the law of gravity. Newton realized that the Moon is not immune to gravity and is continuously falling towards the Earth, but it keeps missing it! Newton thus realized that gravity was not something special to the Earth, but it also acts in space. This was a profound and revolutionary idea. According to Aristotle, the laws governing the heavens were considered to be completely different from the laws of physics here on Earth. Now, however, if the moon was affected by gravity, then it made sense that the rest of the solar system should also be subjected to gravity. Newton found that he could explain the entire motion of the solar system from the planets to the moons to the comets with a single law of gravity. Newton's Universal Law of Gravity states that `all bodies attract all other bodies, and the strength of the attraction is proportional to the masses of the two bodies and inversely proportional to the square of the distance between the bodies'. This is called \textit{universal} because it applies to all bodies in the universe regardless of their nature (We know that it is not completely ``universal'' because zero-mass objects do not feel gravity in the Newtonian picture and in this sense, apart from many other, Einstein's theory is more universal than Newton's. Of course during the time of Newton, zero-mass object would have made no sense.). A modern mathematical way of saying this is,
\begin{equation}
F=\dfrac{GMm}{R^2}
\end{equation} 
where $G$ is Newton's gravitational constant, $M$ and $m$ are the masses of the objects and $R$ is the distance between the objects. This law can be expressed in differential form as Poisson's equation,
\begin{equation}
\nabla^2 \phi= 4 \pi G \rho
\end{equation} 
where $\phi$ is gravitational potential and $\rho$ is the mass density of the object.

Despite its power in explaining the planetary orbits in the solar system, Newton was unhappy with the lack of a mechanism by which gravity worked. Until then, all forces were believed to be \textit{contact forces} - except the gravity. The Newtonian concept of ``action-at-a-distance" was profoundly disturbing to his opponents who attacked his theory as an ``occult''. 

From the period immediately following Newton's discovery of his Universal Law of Gravitation, to about the turn of the nineteenth century, the theory of gravitation stayed essentially unchanged. More sophisticated mathematical tools for understanding the interplay of the planets were developed, but the underlying theory remained stable. The excitement during this period mainly came from the systematic application of the theory of gravity to the heavens. For example, Halley's prediction of the return of the comet that now bears his name; discovery of the Neptune by John Adams and Urbain Leverrier; William Hershel's observations of binary stars and the calculation of the mass of stars; James Maxwell's (the same Maxwell known in electrodynamics and thermodynamics) explanation of rings of the Saturn. Of course, other advances were made, among the most important were the experiments of Cavendish who directly demonstrated the gravitational force between two objects in the laboratory.

\subsection{Einstein's Gravity}

The twentieth century was a time of tremendous progress in physical science. For the understanding of gravity, the century began with two puzzles. The first of these puzzles concerned the orbit of the planet Mercury. The second puzzle was related to a series of experiments performed by the Hungarian physicist Roland E\"otv\"os at the end of the nineteenth century. E\"otv\"os was intrigued by the curious link between Newton's laws of gravity and motion. His experiments showed that the gravitational mass was the same as the inertial mass to at least a few parts in a hundred million.

Einstein's theory of General Relativity \cite{Einstein}, published in 1915, is our most detailed mathematical theory for how gravity works. The foundation stone for the general relativity is the \textit{equivalence principle}, which assumes equivalence between the inertial mass with the gravitational mass. This implies `the weak equivalence principle', \textit{i.e.,} the effects of gravitation can be transformed away locally by using suitably accelerated frames of reference. This can be generalized to `the strong equivalence principle', which allows us to study gravitational interaction by studying only the geometry of the space-time. The modern approach to gravity as the geometry of curved spacetime is based on this theory. 

To understand the geometry of spacetime, consider the distance between two spacetime points in any inertial frame,
\begin{equation}
ds^2=c^2 dt^2-dx^2-dy^2-dz^2.
\end{equation} 
But if these two points are not connected by a straight line, the distance can be given by a more general form,
\begin{equation}
ds^2=g_{\mu\nu}dx^\mu dx^\nu,
\end{equation} 
where sum over repeated indexes is implied. The indexes $\mu,\nu=0,1,2,3$ run over four spacetime coordinates. The coefficient $g_{\mu\nu}$ is a function of the spacetime coordinate $x^\mu$. This is called the metric and it specifies the geometry of the spacetime. To study the geometry of any spacetime for an understanding of the theory of gravity, it is enough to study the metric $g_{\mu\nu}$.

It follows then from the Principle of Equivalence that the equations which govern gravitational fields of arbitrary strength must take the form,
\begin{equation}
G_{\mu\nu}= \frac{8\pi G}{c^4} T_{\mu\nu}
\end{equation} 
where $G_{\mu\nu}$ is called the Einstein tensor which has the geometrical information about the spacetime, $G$ is the Newton's gravitational constant and $T_{\mu\nu}$ is the energy-momentum tensor of the matter present. Einstein tensor is given by,
\begin{equation}
G_{\mu\nu}=R_{\mu\nu}-\frac{1}{2}g_{\mu\nu}R
\end{equation} 
where $R_{\mu\nu}$ is the Ricci curvature tensor and $R$ is Ricci scalar.

In general relativity, one performs calculations to compute the evolution and structure of an entire universe at a time. A typical way of working is to propose some particular collection of energy and matter in the universe, to provide the $T_{\mu\nu}$. Given a particular $T_{\mu\nu}$, the Einstein equation turns into a system of second order nonlinear differential equations whose solutions give us the metric of spacetime $g_{\mu\nu}$, which holds all the information about the structure and evolution of a universe with that given $T_{\mu\nu}$.

General Relativity is perhaps the most beautiful physical theory and one of the crowning glories of modern physics. It is powerful, pleasing to the aesthetic sense and well-tested. General Relativity has survived many different tests, and it has made many predictions which have been confirmed. 
The recently concluded experimental investigation using the satellite-based mission Gravity Probe B confirms the two fundamental predictions of general relativity, the geodetic and frame-dragging effects \cite{Everitt:2011hp}. 
The detection of gravitational waves is one of the most fundamental predictions of general relativity which has not been confirmed as of today. Currently many state-of-the-art gravitational wave detectors are in operation. However none of them have the sensitivity to directly detect the gravitational waves yet \cite{GWreview}.
 Other tests focus on the laboratory-scale measurements to look for signs of extra dimensions, such as a deviation from inverse-square \cite{Kapner:2006si} and missing energy signals in CMS \cite{Chatrchyan:2011jx} and ATLAS \cite{Franceschini:2011wr} experiments at the Large Hadron Collider of CERN. Data from any of these experimental studies will greatly improve our understanding of gravity, and will show us how to go beyond the mathematics of General Relativity to create an even-better theory. 

The unification of quantum theory and general relativity has been a major problem in physics ever since these theories were proposed. The problem is that since all fields carrying energy are affected by gravity, gravity contributes to its own source. When trying to do calculations on the energy scale where gravity is usually thought to be similar in strength as the other forces, the graviton self coupling causes fluctuations which introduces infinities in the calculations. This has lead many theories to accommodate the idea of extra dimensions to get the quantum gravity. A recent non-technical review of extra dimensional theories can be found in \cite{VHS-PKS1}. One of the early possibilities for such a unification of the then known interactions i.e., gravity and electromagnetism, was suggested by Kaluza \cite{Kaluza} and Klein \cite{Klein}. But historically, it was Gunnar Nordstr\"om \cite{Nordstrom} who brought the idea of extra spacial dimension into physics.

\section{Kaluza-Klein Theory}
An early proposal to unite general relativity  and classical electrodynamics was given by Theodor Kaluza \cite{Kaluza} in 1921. He showed that the gravitational and electromagnetic fields stem from a single universal tensor and such an intimate combination of the two interactions is possible in principle, with the introduction of an additional spacial dimension. Although our rich physical experience obtained so far provides little suggestion of such a new spacial dimension, we are certainly free to consider our world to be four dimensional spacetime of the bigger five dimensional spacetime. In this scenario, one has to take into account the fact that we are only aware of the spacetime variation of state-quantities, by making their derivatives with respect to the new parameter vanish or by considering them to be small as they are of higher order. This assumption is known as the \textit{cylindrical condition}. 

The five dimensional line element is given by 
\begin{equation}
d\hat{s}^2 = \hat{g}_{\hat{\mu}\hat{\nu}}(x^\mu,y)d\hat{x}^{\hat{\mu}} d\hat{x}^{\hat{\nu}}
\end{equation} 
with $y$ as the additional spatial coordinate. The five dimensional metric can be expressed as,
\begin{equation}
\hat{g}_{\hat{\mu}\hat{\nu}}=
\left( \begin{array}{cc}
g_{\mu\nu} & g_{\mu 5}\\
g_{5 \nu} & g_{5 5}\\
\end{array}\right)
\end{equation} 
 where all unhatted quantities are four-dimensional and all hatted quantities are five-dimensional.
 
Once we have a spacetime metric, like in standard general relativity we can construct the Christoffel symbols $\Gamma^\mu_{\nu \rho}$, the Riemann-Christoffel curvature tensor $R^\mu_{\nu \rho \sigma}$, the Ricci tensor $R_{\mu \nu}$, the curvature invariant $R$ and then the field equations.
This approach gave a striking result, the fifteen higher-dimensional field equations naturally broke into a set of ten formulae governing a  tensor field representing gravity, four describing a vector field representing electromagnetism, and one wave equation for a scalar field.  Furthermore, if the scalar field was constant,  the vector field equations were just Maxwell's equations in vacuo, and the tensor field equations were the 4-dimensional Einstein field equations sourced by an electromagnetic field. 

In one fell swoop, Kaluza had written down a single covariant field theory in five dimensions that yielded the four dimensional theories of general relativity and electromagnetism! But many problems plagued Kaluza's theory. Not the least of which was the nature of the fifth dimension. There was no explanation given for Kaluza's \textit{ad hoc} assumption, the cylindrical condition.

In 1926, Oscar Klein \cite{Klein} provided an explanation for Kaluza's fifth dimension by proposing it to have a circular topology so that the coordinate $y$ is periodic i.e., $0 \leq y \leq 2\pi R,$  where $R$ is the radius of the circle $S^1$. Thus the global space has topology $R^4 \times S^1$.  So Klein suggested that there is a little circle at each point in four-dimensional spacetime. This is the basic idea of Kaluza-Klein compactification. Although there are four space dimensions, one of the space dimensions is compact with a small radius. As a result, in all experiments we could see effects of only three spacial dimensions. Thus Klein made the Kaluza's fifth dimension less artificial by suggesting plausible physical basis for it in compactification of the fifth dimension. The theory of gravity on a compact space-time is called \textit{Kaluza-Klein theory}. A detailed pedagogical account of this is given in the reference \cite{Duff}.

We introduce the following notations,
\begin{eqnarray}
\hat{g}_{5 5}&=& \phi\\
\hat{g}_{5 \mu}&=& \kappa \phi A_\mu\\
\hat{g}_{\mu \nu}&=&g_{\mu \nu}+  \kappa^2 \phi A_\mu A_\nu.
\end{eqnarray} 
Hereby the quantities $\hat{g}_{\hat{\mu} \hat{\nu}}$ are redused to known quantities. Now, the new metric can be written as 
\begin{equation}
\hat{g}_{\hat{\mu}\hat{\nu}}=\phi^{-\frac{1}{3}}
\left( \begin{array}{cc}
g_{\mu\nu}+ \kappa^2 \phi A_\mu A_\nu & \kappa \phi A_\mu\\
\kappa \phi A_\nu & \phi\\
\end{array}\right),
\end{equation}
where the field $\phi$ appears as a scaling parameter in the fifth dimension and is called the dilaton field. The fields $g_{\mu\nu}(x,y), A_\mu(x,y)$ and $\phi(x,y)$ transform respectively as a tensor, a vector and a scalar under four-dimensional general coordinate transformations.

The Einstein-Hilbert action for five dimensional gravity can be written as,
\begin{equation}
{\hat S}= \frac{1}{2\hat{k}^2}\int d^5\hat{x} \sqrt{-\hat{g}}\hat{R}
\end{equation} 
where $\hat{k}$ is the five dimensional coupling constant and $\hat{R}$ is the five dimensional curvature invariant. We can get the field equations of gravity and electromagnetism from the above action by variational principle.

As Klein suggested, the extra dimension has become compact and satisfies the boundary condition
\begin{equation}
y=y+2\pi R,
\end{equation} 
all the fields are periodic in $y$ and may be expanded in a Fourier series,
\begin{eqnarray}
g_{\mu\nu}(x,y)&=& \sum_{n=-\infty}^{+\infty}g_{\mu\nu n}(x) e^{in\cdot y/R}\\
A_{\mu}(x,y)&=& \sum_{n=-\infty}^{+\infty}A_{\mu n}(x) e^{in\cdot y/R}\\
\phi(x,y)&=& \sum_{n=-\infty}^{+\infty}\phi_n(x) e^{in\cdot y/R}
\end{eqnarray} 

The equations of motion corresponding to the above action are,
\begin{eqnarray}
(\partial^\mu\partial_\mu - \partial^y\partial_y) g_{\mu\nu}(x,y)&=&(\partial^\mu\partial_\mu + \dfrac{n^2}{R^2}) g_{\mu\nu n}(x)= 0\\
(\partial^\mu\partial_\mu - \partial^y\partial_y) A_{\mu}(x,y)&=&(\partial^\mu\partial_\mu + \dfrac{n^2}{R^2}) A_{\mu n}(x) = 0\\
(\partial^\mu\partial_\mu - \partial^y\partial_y) \phi(x,y)&=&(\partial^\mu\partial_\mu + \dfrac{n^2}{R^2}) \phi_n(x) \;\; = 0
\end{eqnarray} 
Comparing these with the standard Klein-Gordon equation, we can say that only the zero modes $(n=0)$ will be massless and observable at our present energy and all the excited states, called as \textit{Kaluza-Klein states,} will have masses 
\begin{equation}
m_n\sim \dfrac{|n|}{R}
\end{equation} 
as well as charge
\begin{equation}
q_n=\sqrt{2}\kappa \dfrac{n}{R}
\end{equation}  
as shown by Salam and Strathdee \cite{Salam}, where $n$ is the mode of excitation. So, in four dimensions we shall see all these excited states with mass or momentum $\sim O(n/R)$.  Since we want to unify the electromagnetic interactions with gravity, the natural radius of compactification will be the Planck length,
\begin{equation}
R=\dfrac{1}{M_p}
\end{equation} 
where the Planck mass $M_p \sim 10^{19}\,GeV.$

Since the Kaluza-Klein metric is a $5 \times 5$ symmetric tensor, it has 15 independent components. However, because of various gauge fixings we will have only 5 independent degrees of freedom. Whereas in four dimensions we have only 2 degrees of freedom for a massless graviton. This implies that from four dimensional point of view a higher dimensional graviton will contain particles other than just ordinary four dimensional graviton. The zero-mode of five dimensional graviton contains a four dimensional massless graviton with 2 physical degrees of freedom; a four dimensional massless gauge boson with 2 physical degrees of freedom and a real scalar with 1 physical degree of freedom. Whereas the non-zero mode of five dimensional graviton is massive and has 5 physical degrees of freedom. 

Kaluza and Klein's five dimensional version general relativity, although flawed, is an example of such an attempt to unite the forces of nature under one theory. It led to glaring contradictions with experimental data. But some physicists felt that it was on the right track, that it in fact did not incorporate enough extra dimensions! This led to modified versions of Kaluza-Klein theories incorporating numerous and extremely small extra dimensions. 
The three main different approaches to higher dimensional unification are
\begin{enumerate}
\item Compactified Approach

In this scenario extra dimensions are forbidden for us to experience as they are compactified and are unobservable on presently accessible energy scales. This approach has been successful in many ways and is the dominant paradigm in the higher dimensional unification. This has lead to new theories like 11-dimensional supergravity, 10-dimensional superstring theory, the latest 11-dimensional M-theory and Braneworld theory.

\item Projective Approach

Projective theories were designed to emulate the successes of Kaluza-Klein theory without epistemological burden of a real fifth dimension. In this way of unification, the extra dimensions are treated as mathematical artifacts of a more complicated theory. The fifth dimension is absorbed into ordinary four dimensional spacetime by replacing the classical tensors of general relativity with projective ones, which in turn alters the geometrical foundation of general relativity itself.

\item Noncompactified Approach

This approach prefers to stay with idea that the new coordinates are physical. Following Minkowski's example, one can imagine coordinates of other kinds, scaled by appropriate dimension transporting parameters to give them units of length. In this approach the extra dimensions may not necessarily be spacelike. This takes the observable quantity such as rest mass as the extra dimension.

\end{enumerate} 
Here we discuss only the compactified approaches and the interested readers can refer to \cite{Overduin:1998pn} for the detailed review and the comparative study of these three approaches.

\section{String Theory}

After the discovery of nuclear interactions, physicists found that it no longer seemed that the Kaluza-Klein theory with one extra dimension was a viable candidate to include all the gauge interactions. The electromagnetic interaction could be accomodadted with only one extra dimension. But the strong, weak and electromagnetic interactions, i.e. the $SU(3) \times SU(2) \times U(1)$ gauge theory requires more degrees of freedom than a 5-metric could offer. However, the way in which to address the additional requirements of modern physics is not hard to imagine, one merely has to further increase the dimensionality of theory until all of the desired gauge bosons are accounted for.  Then how many dimensions do we need to unify modern particle physics with gravity via the Kaluza-Klein mechanism? 
The answer comes from N=8 supersymmetry which contains spin-2 particle. When N=8 supersymmetry is coupled with general relativity, one has 11 dimensional supergravity theory \cite{Witten:1981me}. But it was realized that it is not possible to get all the gauge interactions and the required fermion contents of the standard model from this theory \cite{Witten:1981me}. Then there were attempts to consider 11 dimensional theories with gauge groups. Of course, the main motivation of obtaining all gauge interactions and gravity from one Einstein-Hilbert action at 11-dimensions would be lost, but still this became an important study for sometime. In this construction the main problem was due to new inconsistency, the anomaly.

This problem was tackled with String theory \cite{Becker:2007zj}. Briefly, the origin of string theory was the discovery by Veneziano \cite{Veneziano} and Virasoro \cite{Virasoro} of simple formulas as a model for describing the scattering of hadrons. These formulae revealed a rather novel mathematical structure which was soon interpreted by the physical picture based on the relativistic dynamics of strings by Nambu, Nielsen and Susskind \cite{Nambu}. This string interpretation of `dual resonance model' of hadronic physics was not influential in the development of the subject until the appearance of the 1973 paper by Goddard \textit{et al} \cite{Goddard}. It explained in detail how the string action could be quantized in light-cone gauge. Interestingly, among the massless string states, there is one that has spin two. In 1974, it was shown by Scherk and Schwarz \cite{Scherk} and independently by Yoneya \cite{Yoneya} that this particle interacts like graviton, so the theory actually includes general relativity. This lead them to propose that string theory should be used for unification rather than for hadrons. This implied, in particular that the string length scale should be comparable to the Planck length, rather than the size of hadrons i.e., $10^{-15}\,m$, as it was previously assumed. All this made string theory a potential candidate to be a theory of quantum gravity. 

String theory replaces all elementary point-particles that form matter and its interactions with a single extended object of vanishing length. Thus every known elementary particle, such as the electron, quark, photon or neutrino corresponds to a particular vibration mode of the string. The diversity of these particles is due to the different properties of the corresponding string vibrations. In fact the laws of quantum mechanics tell us that a single elementary string has infinite number of vibrational states. Since each such vibrational state behaves as a particular type of elementary particle, string theory seems to contain infinite types of elementary particles. This would be in contradiction with what we observe in nature were it not for the fact that most of these elementary particles in string theory turn out to be very heavy, and not observable in present experiments. Thus there is no immediate conflict between what string theory predicts and what we observe in actual experiments. On the other hand these additional heavy elementary particles are absolutely essential for getting finite answers in string theory.

The possible advantage of string theory is that the anomalies faced by Supergravity are fixed naturally by the extended nature of strings. The analog of a Feynman diagram in string theory is a two-dimensional smooth surface, and the loop integrals over such a smooth surface lack the zero-distance, infinite momentum problems of the integrals over particle loops. In string theory infinite momentum does not even mean zero distance, because for strings, the relationship between distance and momentum is roughly like 
\begin{equation}
\bigtriangleup L \sim \dfrac{\hbar}{p} + \alpha^{\prime} \dfrac{p}{\hbar}
\end{equation} 
The parameter $\alpha^{\prime}$ is related to the string tension, the fundamental parameter of string theory, by the relation 
\begin{equation}
T_{string}=\dfrac{1}{2\pi\alpha^{\prime}}
\end{equation} 
The above relation implies that a minimum observable length for a quantum string theory is
\begin{equation}
L_{min}\sim 2\sqrt{\alpha^{\prime}}
\end{equation} 
Thus zero-distance behavior which is so problematic in quantum field theory becomes irrelevant in string theories, and this makes string theory very attractive as a theory of quantum gravity. 

If string theory is a theory of quantum gravity, then this minimum length scale should be at least the size of the Planck length, which is the length scale made by the combination of Newton's constant, the speed of light and Planck's constant 
\begin{equation}
L_{p}=\sqrt{\dfrac{\hbar G_N}{c^3}}= 1.6\times 10^{-35} m
\end{equation} 

All was well, but this was only consistent if the dimension of spacetime is 26 and had only gauge bosons in it. Moreover these bosonic string theories are all unstable because the lowest excitation mode, or the ground state, is a tachyon. Adding fermions to string theory introduces a new set of negative norm states or ghosts. String theorists learned that all of these bad ghost states decouple from the spectrum when two conditions are satisfied: the number of spacetime dimensions is 10, and theory is supersymmetric, so that there are equal numbers of bosons and fermions in the spectrum. The resulting consistent string theories are called \textit{Superstring} theories and they do not suffer from the tachyon problem that plagues bosonic string theories.

A very nice feature of such superstring theories is that, in 10-dimensions the gauge and gravitational anomalies cancel for $E_8 \times E_8$ group and the $SO(32)$ group. It was then found that when the extra six-dimensional space is compactified, the four-dimensional world contains all the required fermions and the standard model gauge groups. Supersymmetry could remain unbroken till the electroweak scale to take care of the gauge hierarchy problem. This the appears to be the unified theory of all know interactions. At that time (1984-85), string theorists believed there were five distinct superstring theories. They differ by very general properties of the strings \cite{Witten-AMS}: 
\begin{itemize}
\item{In the first case (\textbf{Type I}) the strings are unoriented and insulating and can have boundaries in which case they carry electric charges on their boundaries.}
\item{In two theories (the \textbf{Type IIA} and \textbf{Type IIB}) the strings are closed and oriented and are electrical insulators.}
\item{In two theories(the \textbf{heterotic} superstrings with gauge group \textbf{SO(32)} and $\mathbf{E_8 \times E_8}$) the strings are closed, oriented and sperconducting.}
\end{itemize} 
But now it is known that this naive picture was wrong, and that the five superstring theories are connected to one another as if they are each a special case of some more fundamental theory. In the mid-nineties it was learned that various string theories floating around were actually  related by duality transformations known as \textit{T-duality} and \textit{S-duality}. T-duality is a symmetry of string theory, relating type IIA and type IIB string theory, and the two heterotic string theories. S-duality relates Type I string theory  to the heterotic SO(32) theory. Using various known dualities between different compactification of different string theories one can now argue that all five string theories are different ways of describing a single theory. These ideas have collectively become known as \textit{M-theory}, where M is for membrane, matrix, or mystery, depending on your point of view! 

In string theory, we assume that the particles we see around us are actually like strings. Since the entire string propagates with time, we have to apply boundary conditions to the end points for consistency. This lead us to either open or closed strings, which have different boundary conditions. When this theory was extended to a membrane (or brane for short), one has to apply boundary conditions to its boundary surfaces. This can then be extended to higher $n$-dimensional branes. In general, branes are static classical solutions in string theories. A $p$-brane denotes a static configuration which extends along $p$-spatial directions and is localized in all other directions. A $p$-brane is described by a $(p+1)$-dimensional gauge field theory. Strings are equivalent to 1-branes, membranes are 2- branes and particles are 0-branes. 

A special class of p-branes in string theory are called D-branes. Roughly speaking, a D-brane is a p-brane where the ends of open strings are localized on the brane. D-branes were discovered by investigating T-duality for open strings. Open strings don't have winding modes around compact dimensions, so one might think that open strings behave like particles in the presence of circular dimensions.

Although these theories now appear to be far from any experiments, it is now established that these theories have the prospect of becoming theory of everything. The scale at which this theory is operational is close to the Planck scale. This makes it experimentally non-viable for a very long time, or probably at any time!

\section{Braneworld Models}

The large separation between the weak scale $(10^3 \,GeV)$ and the traditional scale of quantum gravity, the Planck scale $(10^{19}\, GeV)$ is one of the most puzzling aspects of nature. This is known as the \textit{hierarchy problem}. One theoretical means of solving this problem is to introduce supersymmetry. Alternatively one may hope to address the hierarchy by exploiting the geometry of spacetime.  An extremely popular theory which cures the hierarchy problem by changing the geometry of spacetime with extra space dimensions is the so-called \textit{braneworld} scenario. 

This phenomenological model has been motivated by the work of Horava and Witten \cite{HW}, who found a certain 11-dimensional string theory scenario where the fields of the standard model are confined to a 10-dimensional hypersurface, or brane. In this picture, the  non-gravitational degrees of freedom are represented by strings whose endpoints reside on the brane and on the other hand, gravitational degrees of freedom in string theory are carried by closed strings, which cannot be tied-down to any lower-dimensional object. Hence, the main feature of this model is that the standard model particles are localized on a three dimensional space called the \textit{brane}, while gravity can propagate in $4+n$ dimensions called the \textit{bulk}. It is usually assumed that all $n$ dimensions are transverse to the brane and have a common size $R$. However, the brane can also have smaller extra dimensions associated with it, of size $r \ll R$ leading to effects similar to a small finite thickness. 

The three main features of braneworld models are 
\begin{enumerate}
\item \textbf{Localization of standard model particles on the brane:} A first particle physics application of this idea was put forwarded by Rubakov and Shaposhnikov \cite{Rubakov} and independently by Akama \cite{Akama}. 
\item \textbf{Localization of gauge fields on the brane:} A mechanism for gauge field localization within the field theory context was proposed by Dvali and Shifman \cite{Dvali}. Localization of gauge fields is a rather natural property of D-branes in closed string theories \cite{Polchinski}. 
\item \textbf{Obtaining four-dimensional gravity on the brane:} All the existing braneworld models obtain the laws of (3+1) dimensional gravity on the brane as their low energy approximation.
\end{enumerate} 

The size and geometry of the bulk, as well as the types of particles which are allowed to propagate in the bulk and on the brane, vary between different models. Some important braneworld models are discussed briefly here in the order of their appearance in literature. Somewhat detailed discussion is given in the reference \cite{GG}.

\subsection{Braneworlds with Compact Extra Dimensions}

Here, to obtain  (3+1) dimensional gravity on the brane the idea of KK compactification is combined with braneworld idea. This was proposed in 1998 by Arkani-Hamed, Dimopoulos and Dvali \cite{ADD} along with Antoniadis \cite{AADD}. The additional dimensions are compact, may be as large as as micrometer! As one of its attractive features, the model can explain the \textit{weakness} of gravity relative to the other fundamental forces of nature. In the brane picture, the other three SM interactions are localized on the brane, but gravity has no such constraint and ``leaks" into the bulk. As a consequence, the force of gravity should appear significantly stronger on small say, sub-millimeter scales, where less gravitational force has ``leaked". This opens up new possibilities to solve the Higgs mass hierarchy problem and gives rise to new predictions that can be tested in accelerator \cite{Chatrchyan:2011jx, Franceschini:2011wr}, astrophysical \cite{SN} and table-top experiments \cite{Kapner:2006si}. 

The action for gravity in (4+n) dimensions is given by,
\begin{equation}
S_{4+n}= \dfrac{M_\ast^{2+n}}{2} \int d^4 x \int_0^{2\pi R} d^n y \sqrt{G} {R_{4+n}} + \int d^4 x \sqrt{g} (T+ L_{SM})
\end{equation} 
where $M_\ast \sim (1-10) TeV,$ $g(x)=G(x,y=0)$ and $T+ \langle L_{SM}\rangle=0.$ The low effective four dimensional action for a zero mode takes the form,
\begin{equation}
S= \dfrac{M_\ast^{2+n}{2\pi R}^n}{2} \int d^4 x \sqrt{{g_{zm}}} {R_{zm}}+ \int d^4 x \sqrt{g} (T+ L_{SM}).
\end{equation} 
Comparing it with standard four dimensional pure gravity action we get,
\begin{equation}
M_{pl}^2= M_\ast^{2+n}(2\pi R)^n.
\end{equation} 
Postulating that new quantum gravity scale is at a few TeV, we find the size of the extra dimensions to be,
\begin{equation}
R= 10^{\frac{30}{n}-17} cm.
\end{equation} 
For one extra dimension, $n=1$, the size of extra dimension would be $R\sim10^{13} cm$. This is excluded since it would have modified gravity in solar system scale. For $n=2$ we get $R\sim10^{-2} cm$, which is interesting since it predicts modification of four dimensional laws of gravity at submillimeter scale.

Two static sources on the brane interact with the following non-relativistic gravitational potential
\begin{equation}
V(r)=-G_N m_1 m_2 \sum_{n=-\infty}^{n=+\infty} \vert \Psi_n(y=0) \vert^2 \dfrac{e^{-m_n r}}{r},
\end{equation} 
where $\Psi_n(y=0)$ denotes the wave function of $n^{th}$ KK mode at a position of the brane and $m_n=|n|/R$. If $r\gg L$ from the above expression we get
\begin{equation}
V(r)=\dfrac{-G_N m_1 m_2} {r}.
\end{equation} 
This recovers the conventional four dimensional law of Newtonian dynamics. In the limit $r\ll L$ we get,
\begin{equation}
V(r)=\dfrac{-m_1 m_2} {M_\ast^{2+n} r^{1+n}}.
\end{equation} 
This is the law of (4+n) dimensional gravitational interactions. Therefore, the laws of gravity are modified at distances of order R.

\subsection{Braneworlds with Wrapped Extra Dimensions}

This phenomenon of localizing gravity was discovered by Randall and Sundrum \cite{RS1} in 1999. RS brane-worlds do not rely on compactification to localize gravity on the brane, but on the curvature of the bulk, sometimes called ``warped compactification". What prevents gravity from \textit{leaking} into the extra dimension at low energies is a negative bulk cosmological constant. There are two popular models. The first, called RS-1, has a finite size for the extra dimension with two branes, one at each end. The second, RS-2, is similar to the first, but one brane has been placed infinitely far away, so that there is only one brane left in the model. They also used their model to explain the hierarchy problem \cite{RS2} in particle physics. 

For simplicity, we consider RS-2 model which has a single brane embedded in five dimensional bulk with negative cosmological constant. The action for this model is given by,
\begin{equation}
S_{5}= \dfrac{M_\ast^{3}}{2} \int d^4 x \int_{-\infty}^{+\infty} dy \sqrt{G} (R_{5}-2\Lambda) + \int d^4 x \sqrt{g} (T+ L_{SM}),
\end{equation} 
where $\Lambda$ denotes the negative cosmological constant and $T$ is the brane tension. The equation of motion derived from this action is given by,
\begin{equation}
M_\ast \sqrt{G}(R_{AB}-\frac{1}{2}G_{AB}R)= -M_\ast^{3} \Lambda \sqrt{G} G_{AB} + T \sqrt{g} g_{\mu \nu} \delta_A^\mu \delta_B^\nu \delta(y).
\end{equation} 
In this convention the brane is located in extra space at $y=0$. The above equations have a solution in four dimensional world volume as
\begin{equation}
ds^2=e^{-|y|/R} \eta_{\mu \nu} dx^\mu dx^\nu + dy^2.
\end{equation} 

It is important to emphasize that the five dimensional action is integrable with respect to $y$ for the zero mode. That is, 
\begin{equation}
\dfrac{M_\ast^3}{2} \int d^4 x \int_{-\infty}^{+\infty} dy \sqrt{G} R_{5} \longrightarrow \dfrac{M_\ast^{3}(2 R)}{2} \int d^4 x \sqrt{g} R.
\end{equation} 
The result of this integration is a conventional four dimensional action. Hence we find the relation between four dimensional Planck mass and $M_{\ast}$,
\begin{equation}
M_{pl}^2= M_\ast^{3}(2 R)
\end{equation} 
This looks similar to that in ADD model with one extra dimension. The similarity is due to the fact that the effective size of the extra dimension that is felt by the zero-mode graviton is finite and is of the order of $R$ in both the models.

Besides the zero-mode, there are an infinite number of KK modes. Since the extra dimension is not compactified the KK modes have no mass gap. In the zero mode approximation these states are neglected. However at distances smaller than the size of the extra dimension, the effects of these modes become important.

The static potential between tow sources on the brane is given by 
\begin{equation}
V(r)=\dfrac{-G_N m_1 m_2} {r} \left( 1+\dfrac{(2 R)^2}{r^2}\right).
\end{equation} 
The first term is the conventional four dimensional law of Newtonian dynamics whereas the second term is due to exchange of KK modes which becomes dominant when $r \lesssim R.$

\subsection{Braneworlds with Infinite Volume Extra Dimensions}

This mechanism of obtaining (3+1) gravity on the brane is different from the earlier two as it allows the volume of the extra dimension to be infinite. This model was proposed in 2000 by G. R. Dvali, G. Gabadadze and M. Porrati \cite{DGP}. In the first model the four dimensional gravity could be reprodused at large distances due to finite volume of extra space. This is usually done by compactifying the extra space. Alternatively, this is done by warping the extra dimensions in the second model where still the volume of extra space is finite. But in this scenario the size of the extra dimensions does not need to be stabilized since the extra dimensions are neither compactified nor wrapped because of the presence of infinite-volume extra dimensions and hence gravity is modified at large distances. This gives rise to new solutions for late-time cosmology and acceleration of the universe which comes from type-Ia supernovae observations. This can also explain dark energy problem and Cosmic Microwave Background.

The action in five dimensions with one infinite volume extra dimension is give by, 
\begin{equation}
S_{5}= \dfrac{M_\ast^{3}}{2} \int d^4 x \int_{-\infty}^{+\infty} dy \sqrt{G} R_{5} + \int d^4 x \sqrt{g} (\frac{M_{pl}^2}{2}T+ L_{SM})
\end{equation} 
To study the gravity described by this model, we introduce the quantity
\begin{equation}
r_c \sim M_{pl}^2/M_\ast^3.
\end{equation} 
When $r_c \rightarrow \infty$ the four dimensional term dominates but in the opposite limit $r_c \rightarrow 0,$ the five dimensional term dominates. Therefore we expect that for $r \ll r_c$ to recover the four dimensional laws on the brane while for $r \gg r_c$ five dimensional laws.

The static gravitational potential between the sources in the four dimensional world volume of the brane is given by,
\begin{equation}
V(r)= -\frac{1}{8 \pi^2 M_{pl}^2} \frac{1}{r} \left\lbrace sin\left(\frac{r}{r_c}\right) Ci\left(\frac{r}{r_c}\right)+ 1/2 cos\left(\frac{r}{r_c}\right) \left[\pi-2 Si\left(\frac{r}{r_c}\right)\right]\right\rbrace
\end{equation} 
where $Ci(z)\equiv \gamma + Len(z)+ \int_0^z (cos(t)-1)dt/t$, $Si(z)\equiv \int_0^z (sin(t)dt/t$ and $\gamma \simeq 0.77$ is the Euler-Mascheroni constant, and the distance $r_c$ is defined as follows,
\begin{equation}
r_c \simeq \frac{M_{pl}^2}{2 M_\ast^3}.
\end{equation} 

In this model $r_c$ is assumed do be of the order of the present Hubble size, which is equivalent to the choice $M_\ast \sim 10 - 100\, MeV$. It is useful to study the short distance and the long distance behavior of this expression. At short distance, when $r \ll r_c$ we get,
\begin{equation}
V(r)= -\frac{1}{8 \pi^2 M_{pl}^2} \frac{1}{r}\left\lbrace \frac{\pi}{2} + \left[-1+ \gamma+ ln\left(\frac{r}{r_c}\right)\right] \frac{r}{r_c}+ O(r^2)\right\rbrace
\end{equation} 
Therefore, at short distances the potential has the correct four dimensional Newtonian $1/r$ scaling. This is subsequently modified by the logarithmic `repulsion' term in the above expression. At large distances $r \gg r_c$, the potential takes the form,

\begin{equation}
V(r)= -\frac{1}{8 \pi^2 M_{pl}^2} \frac{1}{r} \left\lbrace \frac{r_c}{r}+ O(r^2)\right\rbrace
\end{equation} 

Thus, the long distance potential scales as $1/r^2$ in accordance with laws of five dimensional theory.

\subsection{Braneworlds with Universal Extra Dimensions}

Universal Extra Dimensions model was proposed by Appelquist, Cheng and Dobrescu \cite{ACD} in 2001. In this model the extra dimensions are accessible to all the standard model fields, referred to here as universal dimensions which may be significantly larger. The key element is the conservation of momentum in the universal dimensions. In the equivalent four-dimensional theory, this implies KK number conservation. In particular there are no vertices involving only one non-zero KK mode, and consequently there are no tree-level contributions to the electroweak observables. Furthermore, non-zero KK modes may be produced at colliders only in groups of two or more. Thus, none of the  known bounds on extra dimensions from single KK production at colliders or from electroweak constraints applies for universal extra dimensions.

The full Lagrangian of this model includes both the bulk and the boundary Lagrangian. The bulk Lagrangian is determined by the SM parameters after an appropriate rescaling. The very important property of this model is the conservation of KK parity that implies the absence of tree level KK contributions to low energy processes taking place at scales very much less than $1/R$ In the effective four dimensional theory, in addition to the ordinary particles of the SM, denoted as zero modes, there are infinite towers of the KK modes. There is one such tower for each SM boson and two for each SM fermion, while there also exist physical neutral and churched scalars with $(n\geq 1)$ that do not have any zero mode partners.

\section{Conclusion}

Many of the major developments in fundamental physics of the past century arose from identifying and overcoming contradictions between existing ideas. For example, the incompatibility of Maxwell's equations and Galilean invariance led Einstein to propose the special	theory of relativity. Similarly, the inconsistency of special relativity with Newtonian gravity led him to develop the general theory of relativity. More recently, the reconciliation of special relativity with quantum mechanics led to the development of quantum field theory. We are now facing another crisis of the same character. Namely, general relativity appears to be incompatible with quantum field theory. Any straight forward attempt to `quantize' general relativity leads to a nonrenormalizable theory. This has lead to theories like superstrings and braneworlds. Even though these theories look rather exotic, at least for the moment. Yet they lead to important insights and also provide a framework for addressing a number of phenomenological issues. Further more, new ideas emerge in approaching fundamental problems which have been puzzling physicists over the centuries.  All this makes the subject interesting and lively. The question is whether the mother nature follows any of these routes being explored in this context?
\section*{Acknowledgement}

The first author (VHS) would like thank Gerald Cleaver and Anzhong Wang for many useful discussions on the topics mentioned in this paper.

\end{document}